# CUPID: Improving Battle Fairness and Position Satisfaction in Online MOBA Games with a Re-matchmaking System

GE FAN*†, Tencent Inc., China

CHAOYUN ZHANG*†, Microsoft, China

KAI WANG†, Bytedance, China

YINGJIE LI, Tencent Inc., China

JUNYANG CHEN‡, Shenzhen University, China

ZENGLIN XU, Harbin Institute of Technology (Shenzhen), China

The multiplayer online battle arena (MOBA) genre has gained significant popularity and economic success, attracting considerable research interest within the Human-Computer Interaction community. Enhancing the gaming experience requires a deep understanding of player behavior, and a crucial aspect of MOBA games is matchmaking, which aims to assemble teams of comparable skill levels. However, existing matchmaking systems often neglect important factors such as players' position preferences and team assignment, resulting in imbalanced matches and reduced player satisfaction. To address these limitations, this paper proposes a novel framework called CUPID, which introduces a novel process called "re-matchmaking" to optimize team and position assignments to improve both fairness and player satisfaction. CUPID incorporates a pre-filtering step to ensure a minimum level of matchmaking quality, followed by a pre-match win-rate prediction model that evaluates the fairness of potential assignments. By simultaneously considering players' position satisfaction and game fairness, CUPID aims to provide an enhanced matchmaking experience.

Extensive experiments were conducted on two large-scale, real-world MOBA datasets to validate the effectiveness of CUPID. The results surpass all existing state-of-the-art baselines, with an average relative improvement of 7.18% in terms of win prediction accuracy. Furthermore, CUPID has been successfully deployed in a popular online mobile MOBA game. The deployment resulted in significant improvements in match fairness and player satisfaction, as evidenced by critical Human-Computer Interaction (HCI) metrics covering usability, accessibility, and engagement, observed through A/B testing. To the best of our knowledge, CUPID is the first re-matchmaking system designed specifically for large-scale MOBA games.

Additional Key Words and Phrases: Matchmaking Systems, Online MOBA Games, Win Prediction, Deep Learning

## 1 INTRODUCTION

Online multiplayer games offer a wealth of data and scenarios that can be leveraged to study user behavior, making them an interactive medium for research. Through the observation and analysis of players' actions, decisions, and reactions, researchers can gain profound insights into user needs, psychology, and behavioral patterns. These insights are highly valuable for the design and evaluation of Human-Computer Interaction (HCI) systems [1–6]. Among these game, Multiplayer Online Battle Arena (MOBA) games have garnered substantial attention in the Computer Supported Cooperative Work (CSCW) community due to their high level of competitiveness and

---

appeal to spectators. They have attracted a large player base and have become a significant source of economic benefits, particularly through events such as E-Sports tournaments that draw substantial audiences. For example, the immensely popular MOBA game *League of Legends* (LoL) boasts a registered player base of 150 million , with over 117 million active players on a monthly basis [7]. Similarly, *Dota 2* as another renowned MOBA game, hosts The International (TI), an annual E-Sports tournament that consistently offers prize pools exceeding tens of millions of dollars to participating teams for several consecutive years [8]. Given the immense popularity and economic significance of MOBA games, understanding players' behaviors within these games has become crucial to enhance their gaming experience. Consequently, substantial attention has been directed towards studying these behaviors and their underlying factors [9–12].

MOBA games usually involve two teams of players compete against each other with the goal of destroying the opposing team's core structure. They are renowned for their amalgamation of real-time strategy, role-playing, and action game elements [13]. Its team-up mechanism relies on the matchmaking system to construct balanced and competitive matches by pairing players of comparable skill levels by searching for suitable teammates and opponents. This aspect is crucial for ensuring a fair, competitive and enjoyable gaming experience. To this end, two key objectives must be fulfilled: *(i)* allowing each player to select their preferred position and character within the game, and *(ii)* maintaining overall game fairness and avoiding overwhelming gameplay[1] [11], wherein a significant skill discrepancy within teams leads to an unsatisfactory experience. The achievement of these objectives is intricately tied to the efficacy of matchmaking systems, as their quality directly impacts players' satisfaction [14, 15], retention [16], and the game's lifecycle [17]. As such, the matchmaking system stands as a pivotal component within the MOBA game.

Existing matchmaking systems in the MOBA games commonly utilize Matchmaking Ratings (MMR) [11, 18–22]. MMR serves as a quantitative measure to assess the skill levels of players, condensing this information into a single-dimensional value. The system calculates the MMR for each player by considering their historical game outcomes and performances. However, the MMR often fail to consider the crucial aspect of team assignment in online matchmaking. In MOBA games, the collective MMR of a team is greatly influenced by the specific assignment of its individual members. Even if players possess the same overall MMR rating, the actual performance of a team can significantly differ due to variations in individual players' proficiency across different positions, roles, or champions. Consequently, there exists a notable disparity between the team's actual skill level and the skill level estimated solely based on MMR. The left section of Figure 1 (subplots (a)(b)(d)) shows a common limitation of conventional MMR-based matchmaking systems. Despite matching players with similar MMR and ensuring that the red and blue teams have equal overall MMR, the absence of a skilled player in the Jungle position on the blue team and wrongly position assignment for player 5 leads to two main issues: *(i)* an imbalanced team assignment, as players are assigned to unfamiliar positions to compensate for the absence, and *(ii)* player dissatisfaction due to being assigned to an undesired position. Consequently, the blue team becomes more vulnerable to defeat against the red team. Moreover, the intricate dynamics between different team assignments introduce coordination and counterplay relationships that further exacerbate this phenomenon.

Furthermore, players in MOBA games often exhibit distinct preferences for different positions, as evident from their pre-selected positions and historical selection frequencies. These position preferences play a vital role in determining the appropriateness of position assignments, individual game proficiency, and overall player satisfaction. However, conventional MMR-based matchmaking systems, which solely rely on a single-dimensional score, often overlook this critical information. Consequently, there is a risk of diminishing the overall gaming experience, leading to lower satisfaction levels and potentially impacting the game's retention rate. Therefore, it becomes imperative for the current matchmaking systems to incorporate this information in order to enhance the players' gaming experience.

---

[1]Overwhelming gameplay refers to a game in which one team's skill level significantly surpasses that of the opposing team, resulting in a substantial mismatch and imbalance. In such games, the outcome is heavily skewed in favor of the more skilled team, leading to an uncompetitive and unsatisfying experience for the less skilled team.

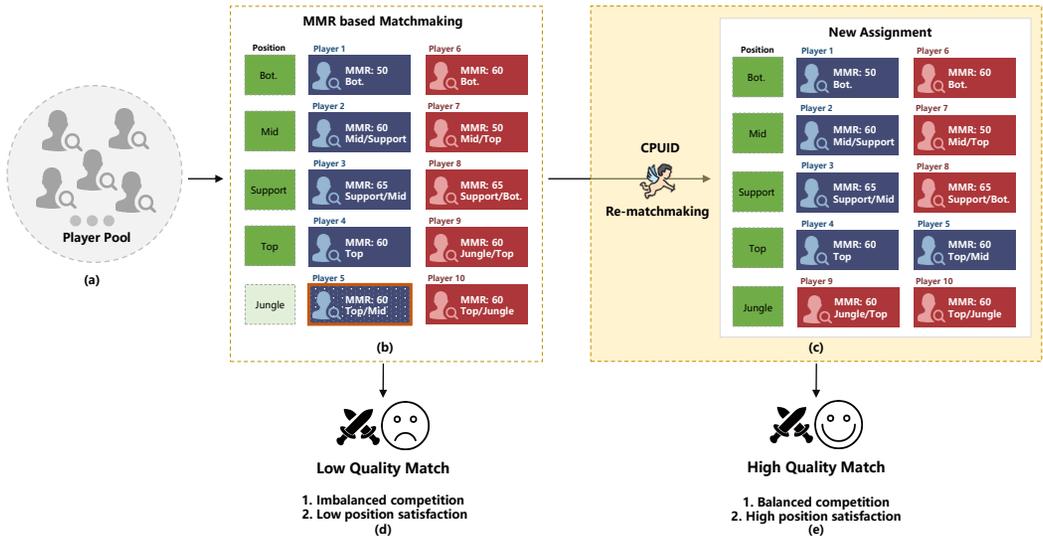

Fig. 1. Overall pipeline of the Cupid matchmaking systems in MOBA games. (a) The full player pool for matchmaking. (b) The MMR-based matchmaking process. The positions assigned by the matchmaking system are displayed on the left side, while the positions listed in the player's block represent their preferred positions. (c) The re-matchmaking system, which optimizes the team and position assignments to maximize fairness and position satisfaction. (d) The matchmaking result, a low quality battle if it fails to ensure fairness and position satisfaction. (e) A high quality battle, which is characterized by appropriate position assignments and a fair battle environment.

Nevertheless, the comprehensive consideration of all the aforementioned factors to determine an appropriate team assignment for a match poses numerous challenges:

(1) **Multi-Objective.** The objective of the comprehensive matchmaking system extends beyond solely relying on MMR. Instead it strives to achieve several primary objectives, including optimizing player position satisfaction and promoting fairness in game matches.

(2) **Skill Heterogeneity.** Ensuring fair team allocation poses a challenge due to the heterogeneity of player skills and the intricate relationships between positions and champions. The outcome of a match depends not only on the individual proficiency of players but also on the synergy between champions and their designated positions, as well as the interactions and counterplays among different champions. Incorporating these diverse factors adds complexity to the task of creating balanced teams.

(3) **Game Uncertainty.** The dynamic nature of MOBA games inherently introduces uncertainty into match outcomes, as it is influenced by a multitude of factors, including strategic decision-making, teamwork, objective control, and adaptability to changing circumstances. These complex interactions introduce a level of uncertainty that poses a challenge when attempting to predict match outcomes solely based on historical data.

(4) **System Efficiency.** Efficiency holds paramount importance in practical industry applications within the realm of matchmaking systems. Players highly value their time and have expectations of prompt matchmaking experiences [14, 23]. Consequently, matchmaking frameworks must strike a delicate balance between accuracy and speed, guaranteeing that players are matched with suitable opponents within a reasonable timeframe.

We propose a matchmaking framework called CUPID[2], which aims to tackle the aforementioned challenges associated with the matchmaking process in MOBA games. Our framework introduces a novel process called "re-matchmaking", which intervenes between the grouping of users with similar MMR ratings to allocate them into different teams and positions. This intervention enhances the fairness of team assignments and improves players' satisfaction with their assigned positions. The re-matchmaking process comprises two key steps that collectively achieve the aforementioned objectives. First, we employ a pre-filtering process that eliminates assignments not meeting players' satisfaction constraints. This step removes a substantial proportion of player assignments, ensuring a minimum threshold of matchmaking quality in terms of position satisfaction while also enhancing overall system efficiency. Second, we design a dedicated pre-match win-rate prediction model called OmniNet within OmniNet (OwO). specifically tailored to the features of players in MOBA games. This model evaluates the fairness of each potential assignment by predicting their win rates. Subsequently, we select the assignment that maximizes the predicted game fairness. This approach further improves game fairness and minimizes the occurrence of overwhelming games. By incorporating the re-matchmaking process, our proposed CUPID framework offers a comprehensive solution to enhance fairness and optimize team assignments in MOBA game matchmaking.

The performance of the CUPID framework is evaluated through extensive offline experiments and online A/B testing conducted in a real environment using a popular mobile MOBA game, encompassing a substantial number of players and matches. The results of both the offline experiments and online A/B testing demonstrate the remarkable performance of the CUPID framework. In summary, the contributions of this paper can be summarized as follows:

(1) **Framework.** We present an innovative re-matchmaking framework named CUPID, aimed at enhancing players' position satisfaction and promoting game fairness in online MOBA games. CUPID incorporates a two-step pre-matchmaking process, which meticulously constructs optimal team assignments and strategically allocates positions. By focusing on both fairness and player satisfaction, the framework ensures an efficient system that significantly enhances the overall game experience for players.

(2) **Model.** In order to achieve precise estimation of game fairness, we introduce a pre-match win prediction model named OmniNet within OmniNet (OwO). OwO is equipped with an ingame economy-aware loss function specifically designed to consider players' profile and performance features from various perspectives. This enables OwO to effectively capture spatial, temporal, and positional information of all players involved in a match, which enhances the accuracy of win prediction and fairness estimation, contributing to an improved assessment of game fairness.

(3) **Analysis.** We perform comprehensive offline experiments utilizing real-world datasets collected from a mobile MOBA game to evaluate the effectiveness of CUPID. The experimental results consistently showcase the superior performance of our method compared to all baseline approaches, as indicated by all evaluation metrics. Notably, our approach achieves a substantial relative improvement in accuracy, surpassing the state-of-the-art methods by on average 7.18%.

---

[2]We employ the name "CUPID" to symbolize our framework's capacity to identify the optimal matching of teammates and opponents for multiplayer game battles, drawing an analogy to the legendary ability of the god of CUPID.

(4) **Player Experience.** We integrate our proposed framework into the matchmaking system of an real mobile MOBA game with millions players basis and perform an online A/B testing specifically on its ranking play mode. The experimental results unequivocally validate the substantial advancements achieved by our framework in terms of both position satisfaction and game fairness in MOBA games, improving critical HCI metrics covering usability, accessibility, and engagement. These improvements notably enhance the overall player experience.

(5) **Efficiency.** Our online deployment demonstrates that CUPID is a highly scalable framework with negligible responsive latency. This indicates that the re-matchmaking process is hardly perceivable by players, preventing any compromise to their matchmaking patience and overall experience.

To the best of our knowledge, the proposed CUPID framework represents the first application of a re-matchmaking process in large-scale MOBA games to enhance player satisfaction and promote fair battles through optimized position and team assignments.

## 2    RELATED WORK

In this section, we will provide an overview of relevant research and industry practices regarding matchmaking systems and the prediction models in online games.

### 2.1    HCI in Online Games

Online games serve as a significant manifestation of the internet era, holding great importance within the CSCW and HCI community [24–28]. Early work in game designs provided macro-level principles and guidelines for games [29, 30]. For instance, research by Campbell *et al.*, underscores the importance of fairness in game design, asserting that all players should have an equal chance of winning [30]. This principle aligns with the core optimization focus of CUPID. Recent research has focused on more detailed studies within specific subdomains [31–33]. Within these works, Kim *et al.*, [34] demonstrate the significance of player proficiency and equivalent team skill as crucial factors in MOBA, significantly impacting team performance. Jiang *et al.*, [35] illustrate differences in players' preferences for positions and champions, emphasizing that playing style profoundly influences their contributions to teamwork. These findings align with the design principles of our re-matchmaking approach. Additionally, research such as [36–39] has investigated the interaction between players and AI. Studies such as [31–33, 40, 41] have focused on interaction design in different scenarios, such as VR and full-body interactions. These research represent important aspects of HCI on interaction design in the context of online games.

The analysis of user experience stands as another hotspot in research [42–47], especially in the realm of MOBA games [48–50]. These insights gained not only enhance player experiences but also contribute to game design, player engagement, and company revenue generation. For example, Kordyaka *et al.*, [49] demonstrate that experiencing negative events is one of the factors contributing to the manifestation of toxic behaviors, such as Away From Keyboard (AFK) incidents. These behaviors are infectious to other players, thereby impacting the overall player experience [44]. The quality of matchmaking can significantly influence the occurrence of these toxic behaviors.

Recent years have witnessed a growing interest in leveraging data-driven techniques to analyze user behavior for personalization, such as data visualization [50–52] and machine learning. These data learning-driven methodologies present new avenues for comprehending and studying personalized user behavior in MOBA games [53], empowering to gain deeper insights and make informed decisions. For instance, Xie *et al.*, [52] proposed RoleSeer, a visual analytics system, which explores informal roles via behavioral interactions, illustrating their dynamic conversions and transitions.

On the other hand, players' gaming experiences are influenced by various factors, encompassing gaming elements [54], in-game items [55], negative behaviors [44], and playtime duration [56]. This underscores the necessity of incorporating player personalization and addressing heterogeneity when designing a matchmaking system with data-driven methods. This concern is effectively optimized by Cupid.

While existing methods predominantly focus on the offline analysis of player behavior in online games, many are constrained to such analyses and lack practical implementation in a live online gaming environment. Cupid stands out as a representative HCI system that addresses this gap, having been deployed in a real large-scale MOBA game.

## 2.2 Matchmaking Systems in Online Games

Various matchmaking systems in online games rely on precise player skill ratings, typically represented as Matchmaking Rating (MMR). Several MMR methods have been proposed in the literature, including ELO [57], ELO-MMR [22], Glicko [18], TrueSkill [19], and their respective variants and extensions [20, 21, 58]. Originally designed for 1 vs. 1 battles, such as Chess and GO, ELO has also been applied to online games [57]. The Glicko rating system offers a more dynamic estimation of player MMR by considering the uncertainty of player ratings and adjusting them based on match outcomes and expected performance [18, 20]. TrueSkill and TrueSkill2, on the other hand, are specifically tailored for team-based games and provide accurate and reliable skill ratings by considering match results and the relative skill levels of players within teams [19, 21]. The Factor-Based Context-Sensitive Skill Rating System (FBCS-SRS) employs a factor model that represents individual MMR as the inner product of an agent factor vector and a context factor vector [59]. In contrast, QuickSkill is a framework that addresses the cold-start problem commonly encountered by traditional MMR models. It achieves this by accurately estimating player skill ratings even in scenarios where limited or no historical data is available [11].

Recent research has extended the scope of game matchmaking systems beyond traditional player skill ratings to incorporate additional factors, such as player behavior, preferences, roles, playstyles, and champion interactions [14, 60]. Veron *et al.*, [14] conducted an analysis of gameplay session data to gain insights into player behavior, preferences, and skill levels. They proposed matchmaking algorithm enhancements aimed at creating more balanced and enjoyable matches for players. Delalleau *et al.*, [61] further expanded on traditional skill rating systems by considering factors such as player roles, playstyles, and preferences. This approach seeks to create more dynamic and strategic matches that account for the diverse characteristics of players. OptMatch [17] explores the intricacies of high-order interactions and dynamics that arise during gameplay. The study focuses on understanding and modeling various relationships in online games, including those between players and champions. In contrast, GloMatch [62] introduces a novel data-driven framework to tackle the globally optimized matchmaking problem. It employs an effective policy-based deep reinforcement learning algorithm to optimize the matchmaking process and enhance overall matchmaking quality.

Most of existing research on game matchmaking systems predominantly employs one-step MMR-based matchmaking approaches, neglecting the diversity in players' skills and proficiency across different positions. Consequently, this limitation in accounting for heterogeneity compromises the fairness and satisfaction levels of the matchmaking system.

## 2.3 Prediction Model in Online Games

Machine learning-based prediction models have gained attentions in the many different fields [63–71], and its popularity in gaming is increasing, with applications ranging from win rate prediction to churn and player behavior analysis [72–78]. With the increasing prominence of esports competitions, the accuracy of win prediction models has become paramount. In general, these models can be broadly categorized into two groups: in-game and pre-match.

In the realm of esports competitions, in-game prediction models have gained significant prominence. These models leverage real-time game state analysis to forecast the likelihood of one team prevailing over the other, taking into account factors such as team assignment, map control, and objective control. In-game prediction models serve a crucial role in various applications, including live commentary and analysis provided by broadcasters during matches and assisting coaches in making strategic decisions during gameplay [12, 79, 80]. NeuralAC is an example of a machine learning model that employs neural networks to capture the effects of cooperation and competition among players, enabling accurate predictions of match outcomes [81]. Another notable model, the Two-Stage Spatial-Temporal Network (TSSTN), has been proposed to forecast the outcomes of Honor of Kings matches in real-time [12]. An intriguing aspect of this model is its ability to attribute the final prediction results to the contributions of various features, ensuring interpretability [82] and understanding of the underlying factors influencing the outcome.

On the other hand, pre-match prediction models are employed to analyze player and team data prior to the start of a game, enabling the prediction of which side is more likely to emerge victorious. These models incorporate various factors, including player statistics, team performance history, and match-ups, to forecast the outcome of a game before it commences. Extensive research has been conducted in this domain for multiple games, such as Dota 2, League of Legends (LoL), and other esports titles. Semenov *et al.*, [72] investigated the utilization of machine learning models to predict the outcome of Dota 2 matches based on the draft phase. Gong *et al.*, [17] employed graph embedding techniques to capture higher-order interactions between champions and players, enabling predictions of match outcomes. Do *et al.*, [83] focused on utilizing neural networks to forecast game outcomes in *LoL* based on players' experience with specific champions. In another study, Hitar *et al.*, [80] explored the effectiveness of various features, including player performance statistics, team assignment, and in-game events, in predicting match outcomes. They also conducted a comparative analysis of several machine learning models, demonstrating that an ensemble learning model employing carefully designed artificial features achieved state-of-the-art performance.

The Cupid framework proposed in this study introduces a novel approach that utilizes pre-match prediction models to enhance game fairness and optimize players' position satisfaction. By accurately predicting the game outcome prior to its commencement, the framework facilitates adjustments in team assignment, resulting in more balanced matches. Notably, this approach incorporates not only player MMR but also considers a range of additional factors that influence the game's ultimate result.

## 3 PRELIMINARY

This section presents an overview of Multiplayer Online Battle Arena (MOBA) games and their matchmaking systems. We also discuss the primary objective and function of the re-matchmaking system.

### 3.1 Background

*3.1.1 MOBA Game.* The Multiplayer Online Battle Arena (MOBA) [13] genre has witnessed a significant surge in popularity within the competitive video game landscape. We first explain the common terms used in MOBA games in Table 1. In a typical MOBA game, two teams, commonly known as the blue team and the red team, engage in virtual battles within an arena setting. Within the classical MOBA game setup, players are matched with teammates and opponents via a matchmaking system. This process results in the formation of teams comprising five players, each controlling a unique game character, who then compete against another team. The primary objective of the matchmaking system is to establish equitable matches by pairing players with comparable gaming abilities. Throughout gameplay, each player selects a champion and collaborates closely with their team to achieve victory. This involves the strategic elimination of enemy characters, destruction of their structures, and ultimately, the annihilation of the opposing team's core structure, often referred to as the "Nexus" or "Ancient". Figure 2(a) showcases a screenshot example of a game in the popular multiplayer online battle arena game *League of Legends* (LoL). The screenshot captures an intense battle between a player-controlled champion and an enemy champion in the bottom lane. This depiction exemplifies the dynamic and action-packed nature of battles that occur within the game.

Table 1. Explanation of common terms used in MOBA games in this paper.

| Term | Explanation |
| --- | --- |
| **Player** | A player refers to an individual who actively participates in the game. Each player controls a champion and works towards achieving the objectives of the game. |
| **Position** | Position refers to the specific role or lane that a player assumes during the game. In MOBA games, such as LoL, common positions include top, jungle, mid lane, bottom and support. |
| **Champion** | Champions are the playable characters in MOBA games. Each champion possesses unique abilities, strengths, and weaknesses. Players select a champion at the beginning of the game and utilize their skills to contribute to their team's success. Champions can be customized with items and can level up throughout the match. |
| **Room** | Room refers to a game lobby or a virtual space where players gather before a match to discuss strategies or make preparations. Players in the same room will be assigned to the same team by the matchmaking system for cooperation. |
| **Team** | A team consists of a group of players who collaborate and work together towards a common goal. In MOBA games, teams typically consist of multiple players who compete against another team. |
| **Battle/Game** | A battle or game refers to the combat or engagement between two teams in the game. It involves using abilities, tactics, and teamwork to defeat opponents and achieve victory. |

To facilitate efficient collaboration, players in MOBA games are typically assigned specific positions within their teams. These positions, often referred to as lanetypes or roles, define the distinct responsibilities and tasks that players undertake during gameplay. While position assignment methods may vary across different MOBA games, positions tend to remain relatively stable within a specific game duration. For example, in *LoL*, positions are categorized as Top, Jungle, Mid, Bottom, and Support. Figure 2(b) illustrates the different lanes on a mini-map in the game, representing the assigned positions for players. The mini-map provides a visual representation of the game map, highlighting specific areas where players fulfill their roles and engage in gameplay activities. Throughout this research, we adopt this standardized naming convention to maintain clarity and consistency.

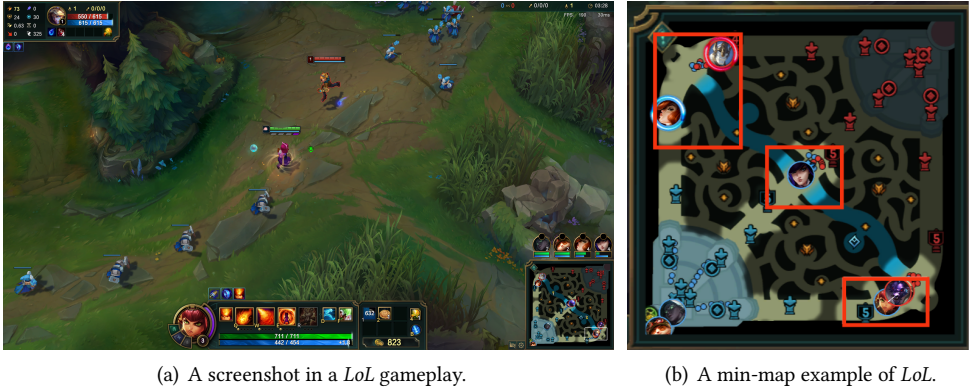

(a) A screenshot in a *LoL* gameplay.                (b) A min-map example of *LoL*.

Fig. 2. Screenshot examples of gamplay in an *LoL* game.

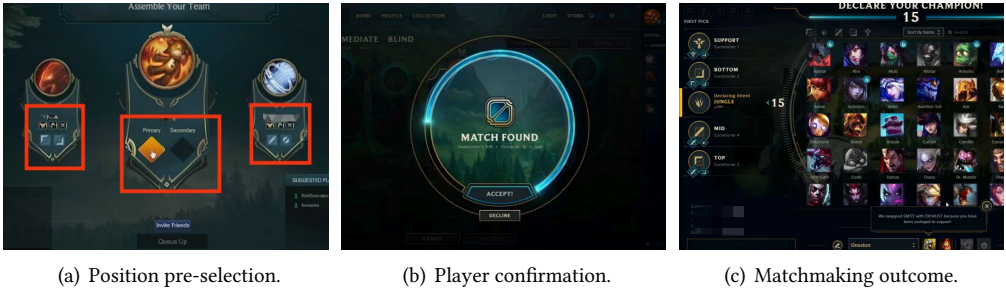

(a) Position pre-selection.        (b) Player confirmation.        (c) Matchmaking outcome.

Fig. 3. An illustrative example of the matchmaking process in *League of Legends* as perceived by a player. (a) Prior to the commencement of the game, players engage in pre-selection of their preferred positions within the game. (b) Once the matchmaking system successfully composes a group of 10 players, each player is prompted to confirm their intention to proceed with the game. (c) With the game officially initiated, players are assigned to their respective positions and teams, and subsequently proceed to choose their individual champions.

*3.1.2  Matchmaking systems in MOBA games.* Matchmaking systems play a crucial role in MOBA games by uniting players of similar skill levels, creating balanced teams, and ensuring fairness and competitiveness. MMR-based matchmaking systems have been widely implemented in the industry for most MOBA games, where each player is assigned a numerical MMR score representing their skill level. This score is determined based on factors such as game outcomes and prior performance. Winning increases MMR, while losing decreases it. During the match creation process, the matchmaking system considers the MMR scores of players to form teams with comparable total MMR, aiming for well-balanced matchups.

In addition to skill-based matchmaking, the allocation of players' positions has gained recognition in recent advancements of industrial game matchmaking systems. These systems enforce players to select pre-defined positions before the game starts. Figure 3 provides an illustrative example of the matchmaking process in *LoL* from a player's perspective. Positions in a MOBA game correspond to distinct lanes on the game map and determine a player's role (e.g., Mid, Support). The matchmaking system takes into account players' position preferences to avoid conflicts and pairs them accordingly. However, this approach imposes rigid position constraints, reducing the number of eligible players in the matchmaking pool. Consequently, the matchmaking process may be prolonged, potentially compromising the overall matchmaking quality.

## 3.2 Re-matchmaking Process Formulation

The objective of the proposed re-matchmaking process in this paper is to optimize player position satisfaction and game fairness by allocating players to teams and positions. This process is illustrated in Figure 1. Initially, the MMR-based matchmaking system gathers 10 players using a MMR-based matchmaking algorithm to form a match. The system then generates multiple sets of position assignments based on satisfaction and fairness metrics. From these assignments, an optimal assignment is selected that maximizes overall position satisfaction and game fairness.

*3.2.1 Team and Position Assignment.* The re-matchmaking process in Cupid aims to achieve the desired outcome by assigning each player to a specific team and position within the game. We represent the assignment of player $n$ to team $t_n$ and position $i_n$ as $A_t^{i,n}$. The primary objective of the team and position assignment is to identify the optimal assignment $\mathcal{A}$ for all players, denoted as $\mathcal{A} = \{A_{t_1}^{i_1,1}, A_{t_2}^{i_2,2}, \cdots, A_{t_N}^{i_N,N}\}$, that maximizes position satisfaction and game fairness. This assignment process forms the fundamental basis of the re-matchmaking process as well as the Cupid framework.

*3.2.2 Position Satisfaction.* Enabling players to select their desired positions is a primary objective of MOBA matchmaking systems. The preference information regarding player positions is explicit and valuable, and it holds significance throughout the entire matchmaking process. This information is scarce and precious, and when a player is unable to obtain their preferred position, it can negatively impact the gaming experience for all participants in the match.

We employ a modeling approach that utilizes the user's historical information and pre-selected position data to determine a position satisfaction score ranging from 0 to 1. This score represents the likelihood of the user being satisfied with a particular position. Higher satisfaction scores indicate a greater level of contentment with the corresponding positions. We assume that the satisfaction probabilities for each user and position are independent. Denoting the satisfaction score of player $n$ for his assigned position $i_n$ as $p_n^{i_n}$, the total satisfaction probability $\mathcal{P}$ of all players for this assignment $\mathcal{A}$ is obtained by the product of individual satisfaction probabilities, *i.e.*,

$$\mathcal{P}_{\mathcal{A}} = \prod_{n=1}^{N} p_n^{i_n}, \tag{1}$$

where $N$ is total number of players.

*3.2.3 Win Rate Prediction.* Win rate prediction is a crucial task in assessing match fairness, as it provides an estimation of the probability that one team will win against another for a specific assignment $\mathcal{A}$ prior to the battle. The predicted win rate serves as a valuable metric for evaluating the fairness of a match. The closer the win rate is to 50%, the higher the level of fairness. To quantify match fairness for a specific assignment $\mathcal{A}$, we utilize the win rate prediction model to transform the win rate prediction results for the assignment $\mathcal{A}$, denoted as $\hat{\mathbf{y}}_{\mathcal{A}}$, into match fairness scores, denoted as $s_{\mathcal{A}}$. The match fairness score is calculated using the following formula:

$$s_{\mathcal{A}} = 1 - 2|\hat{\mathbf{y}}_{\mathcal{A}} - 0.5|. \tag{2}$$

By applying this formula, the match fairness scores scales from 0 to 1, with higher scores indicating a fairer match.

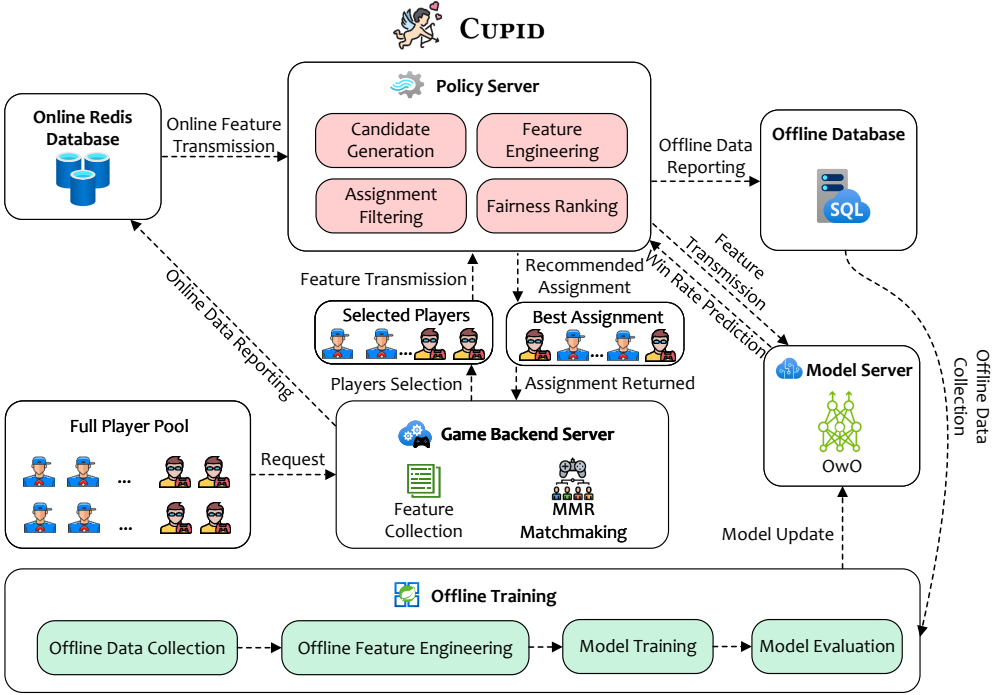

Fig. 4. The overall architecture of Cupid.

## 4 THE DESIGN OF CUPID

In Section 4.1, we present an overview of the overall architecture of Cupid, highlighting its key components and their interactions. In the following subsections, we delve into the details of each selected component, providing a comprehensive understanding of their functionalities and contributions to the overall system.

### 4.1 Cupid in a Nutshell

The overall architecture of Cupid is depicted in Figure 4. It comprises five essential components that collectively support the re-matchmaking process and ensure high-quality matchmaking. These components are the game backend server, policy server, model server, online/offline database systems, and offline training pipeline.

To recap the re-matchmaking process illustrated in Figure 1, when a matchmaking request is received, the game backend server quickly assembles 10 players with similar MMRs using a matchmaking algorithm and collects their corresponding features. These features are then sent to the policy server and stored in an online Redis database for backup. Concurrently, the policy server queries the online database for historical features of the players and filters out assignments that do not meet the satisfaction constraint, generating a set of candidate assignments.

After performing feature engineering, the processed features are sent to the model server, where the OmniNet within OmniNet (OwO) model is employed for win rate prediction. The predicted results are translated into fairness scores, as defined in Section 3.2.3, and sent back to the policy server. The policy server then ranks the assignments based on their predicted fairness scores and recommends the best assignment to the game backend server as the final matchmaking result.

Table 2. Examples and statistics of feature sets used in the prediction model of MOBA games.

| Group Name | Feature Name | Type | Dimemsionality |
|---|---|---|---|
| Team Feature | Team_MMR_Mean | Numerical | $\sim 10^2$ |
| | Team_MMR_Variance | Numerical | $\sim 10^2$ |
| | ... | ... | ... |
| User Short-term Feature | ChampionID | Categorical | $\sim 10^2$ |
| | Kill_Num | Numerical | $\sim 10$ |
| | Result | Binary | 2 |
| | ... | ... | ... |
| User Long-term Feature | Win_Rate | Numerical | $\sim 10$ |
| | Championsid_Top_1 | Categorical | $\sim 10^2$ |
| | ... | ... | ... |
| User Real-time Feature | Normal_MMR | Numerical | $\sim 10^2$ |
| | Is_Promotion | Binary | 2 |
| | Avilibile_position_1 | Categorical | 5 |
| | ... | ... | ... |

Simultaneously, the processed features are logged into an offline database to support offline model training. Regular offline model training is conducted to capture online data drift. After retraining OwO with the latest data and conducting thorough evaluation, the new model is pushed to the model server, replacing the old model.

Overall, the key of Cupid lies on the two processes in the re-matchmaking. Namely,

- **Assignment Filtering Process**, filters out assignments that fall below the position satisfaction threshold and minimum acceptable level of position satisfaction for all players;
- **Win Rate Prediction Model**, which accurately assesses game fairness to ensure balanced battles and minimize overwhelming gameplay.

The following subsections provide detailed descriptions of all key components of Cupid.

## 4.2 Feature Collection

Numerous prior studies have established a positive correlation between model performance and data quality [80, 84, 85]. To accurately characterize MOBA game players, the game backend sever collects a diverse and comprehensive set of features from multiple sources, categorized into different feature groups. These feature groups enable us to comprehensively profile the skill, preferences, behaviors, and habits of the players [86]. The feature groups include:

- **Short-term Features**. These features encapsulate the detailed game performance of the player in the last $K$ games. They include metrics such as the number of kills, deaths, assists, selected positions, and champions. The aim of these features is to capture the player's recent performance and preferences, which reflect their behavioral trends.
- **Long-term Features**. These features encompass the overall statistics of the player in the last 30 games. They include metrics such as the average number of kills and deaths, the number of selections for each position, and the corresponding performance, as well as the most frequently selected champions. The purpose of these features is to provide a high-level overview of the player's performance and preferences.

- **Real-time Features**. These features include the instantaneous MMR scores of the player across different versions of the game, the player's pre-selected position, and their instantaneous position satisfaction. These features capture the player's real-time game skill and preferences.
- **Team-term Features**. These features provide aggregated summaries of both teams in the game, such as the average MMR of each team. These features offer an evaluation at the team level, enabling a broader perspective on the game dynamics.

Table 2 provides examples and statistics of the feature sets utilized in the MOBA game prediction model. The features cover a range of information, including team-related metrics (*e.g.*, Team_MMR_Mean), champion usage (*e.g.*, Championid_Top_1), and player preferences (*e.g.*, Available_position_1). These features capture various aspects of the game and the player's behavior. The original data from the game consists of different types, such as categorical and numerical features. To enhance the model's robustness, we apply feature engineering techniques, such as bucketing, to transform the original features into one-hot encoded representations before feeding them into the model [47, 87]. The "Dimensionality" column in the table indicates the size of the one-hot encoded features, which serve as inputs to the model. By utilizing these diverse features, Cupid is able to comprehensively profile players from different perspectives, surpassing the limitations of MMR-based models that rely solely on a single score. This approach enables a more nuanced understanding of players' characteristics and preferences, leading to more accurate matchmaking and improved gaming experience.

### 4.3 Candidate Assignments Generation and Filtering

Candidate assignment generation and filtering are essential functions performed by the policy server to ensure players' position satisfaction and eliminate redundant assignment candidates. As depicted in Figure 3(a), players have the option to pre-select one or more preferred positions before the match begins, enabling them to showcase their abilities and make a valuable contribution to the team. Pre-selected positions serve as explicit signals, aiming to prevent internal disputes and promote team harmony once the game starts. The satisfaction score for player $n$ in position $i$ is denoted as $p_n^{in}$. For the player's pre-selected position, the satisfaction score is set to the maximum value. For other positions, the satisfaction score is determined based on the player's historical selection frequency. Positions that have been frequently chosen in the past receive higher satisfaction scores, indicating a stronger preference for those positions. This scoring mechanism ensures that players are assigned to positions that align with their preferences, promoting a higher level of satisfaction and a more enjoyable gaming experience.

In a standard MOBA game with 10 players, 2 teams, and 5 positions, the total number of possible assignments $\mathcal{A}$, considering all combinations without any constraints, amounts to $A_{10}^{10} = 3,628,800$ assignments. However, it is impractical to handle such an enormous number of model queries online within the time constraints of a matchmaking request. Therefore, a pre-filtering process is necessary to eliminate redundant assignments that do not align with the objectives of the Cupid framework. To perform this pre-filtering, we leverage the overall satisfaction score $\mathcal{P}_{\mathcal{A}}$ as a filtering criterion. This score serves as one of the key objectives of the re-matchmaking process and can be computed efficiently even with a large number of requests. The satisfaction-based filtering process is both simple and effective. If the overall satisfaction score of an assignment $\mathcal{A}$ falls below a predetermined threshold $\mathcal{P}_\tau$, *i.e.*, $\mathcal{P}_{\mathcal{A}} < \mathcal{P}_\tau$, that assignment is removed from the candidate set. The specific value of $\mathcal{P}_\tau$ is determined by the game designers. Note that in real-world MOBA games, it is common for multiple players to form a group and express a preference to play on the same team before the matchmaking. To accommodate this, our system incorporates a filter mechanism during the assignment process, ensuring that the final assignments align with the desired team composition of players in the same room. This ensures that players within the same group are placed on the same team during the re-matchmaking process.

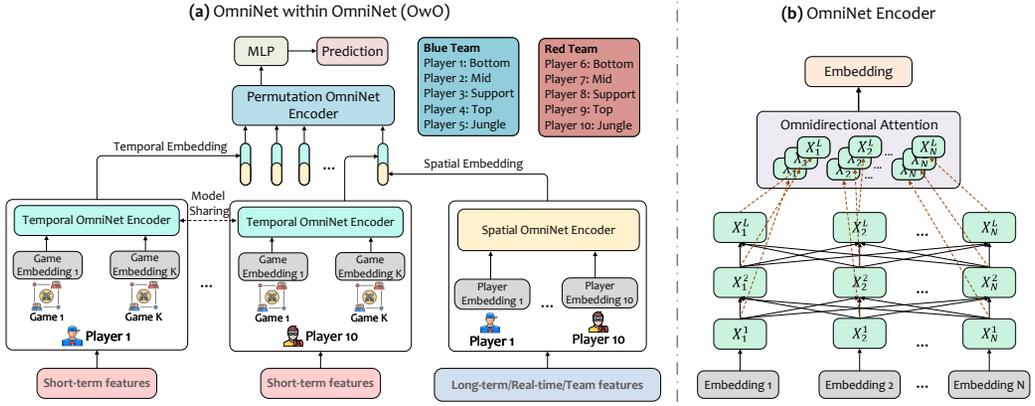

Fig. 5. (a) The overall architecture of the proposed OmniNet within OmniNet (OwO). (b) The structure of the OmniNet encoder.

This pre-filtering step significantly reduces the number of assignments that need to be processed by the fairness prediction model, thereby improving the overall efficiency of the system. Furthermore, it ensures that all assignments retained in the candidate set meet a minimum level of overall position satisfaction, aligning with the primary objective of the re-matchmaking process. The effectiveness and efficiency of this pre-filtering process will be further demonstrated in Section 6.4.

## 4.4 Fairness Assessment

After the satisfaction-based filtering process, the number of candidate assignments is substantially reduced. The remaining assignments are then fed into the win rate prediction model, OmniNet within OmniNet (OwO), which is specifically designed to predict win rates in MOBA games. The goal is to evaluate the fairness of these assignments using the fairness score $s_{\mathcal{A}}$ defined in Section 3.2.3. Among the remaining assignments, Cupid selects the one with the highest predicted fairness as the final matchmaking outcome. Next, we will delve into the design of OwO and explain how we enhance its training using a normalized economy difference loss function. This approach allows us to further optimize the prediction accuracy of win rates, contributing to the overall effectiveness of the re-matchmaking process.

### 4.4.1 OmniNet within OmniNet.
In the OwO model, the primary objective is to predict the win rate of a team based on a given assignment $\mathcal{A}$. The predicted win rate is then translated into a fairness score $s_{\mathcal{A}}$ for further evaluation. Figure 5 (a) provides an overview of the overall architecture of OwO, which consists of several key components, each serving a specific functionality. These components include:

The OwO model plays a pivotal role in the Cupid framework by predicting the win rate of a team based on a given assignment $\mathcal{A}$. The predicted win rate is then translated into a fairness score $s_{\mathcal{A}}$, which serves as a fundamental criterion for evaluating the fairness of different assignments. Figure 5 (a) provides an in-depth view of the overall architecture of OwO, which encompasses three key components, each serving a distinct purpose:

- **Temporal OmniNet Encoder (TOE)**: TOE is responsible for modeling the short-term feature groups of individual players in the last $K$ games. These features capture the recent game performance and preferences of players, enabling the model to incorporate temporal information into the win rate prediction process.
- **Spatial OmniNet Encoder (SOE)**: The SOE component focuses on embedding the long-term, real-time, and team feature groups for each player. By incorporating high-level statistics and information about player behavior and team dynamics, the SOE component provides valuable spatial information that enhances the win rate prediction accuracy.
- **Permutation OmniNet Encoder (POE)**: The POE component aggregates all the processed embeddings from the TOE and SOE components. Its primary role is to capture the win rate differences resulting from the various permutations of the assignment. By considering the different possible combinations of players in a team, the POE ensures that the model accurately predicts the win rate for each assignment.

The output of the POE is subsequently fed into a Multilayer Perceptron (MLP) [88] to perform the final win rate prediction. In our OwO model, we employ individual Temporal OmniNet Encoders (TOEs) to embed the short-term features for each player. These short-term features capture the detailed performance and preferences of a player in the last $K$ games, forming a multivariate sequence. To reduce the overall model complexity, the parameters of the TOEs are shared at the player dimension. This allows us to leverage the collective information from all players while avoiding redundant computations.

Furthermore, in Figure 5 (a), we observe that the input order of the features is fixed at the player level, corresponding to their assigned position and team. For example, the features of the first player input to the OwO model are those of the player assigned to the Bottom position of the blue team. By maintaining this fixed input order, any changes in the assignment, such as a different player being assigned to the Bottom position or a different team composition, will result in a corresponding change in the team features and the input order at the player level. This permutation change is effectively captured by the POE, which is sensitive to variations in the assignment. This ensures that the model is able to accurately predict the win rate for each assignment and account for the impact of player positions and team compositions on game outcomes.

The integration of these three components within the OwO model enables a comprehensive analysis of the assignment, facilitating the generation of accurate win rate predictions. These predictions, in turn, serve as a crucial factor in evaluating the fairness of potential assignments in the re-matchmaking process.

### 4.4.2 OmniNet Encoder.
The OmniNet encoder is a crucial component of the OwO model, consisting of the TOE, SOE, and POE [89]. These encoders build upon the traditional Transformer model [90] and enable the effective encoding of all relevant information pertaining to MOBA players.

We show the structure of the OmniNet Encoder in Figure 5 (b). The Transformer model is widely used for sequential learning tasks, using its multi-head self-attention mechanism [91–96]. OmniNet introduces an additional omnidirectional attention mechanism that enhances the learning capabilities of the model [89]. This attention mechanism connects hidden representations across all layers of the model and automatically weighs their importance. The operation of the omnidirectional attention is defined as follows:

$$O_{attention} = \text{Attention}(X_1^1, X_1^2, \cdots, X_N^{L-1}),  \tag{3}$$

where $O_{\text{attention}}$ is the output of the omnidirectional attention and $X_N^l$ represents the hidden layer in layer $l$ at state $N$. By operating over all hidden layers, the omnidirectional attention produces a tensor of dimensions $L \times T \times d$. The output of the omnidirectional attention, $O_{\text{attention}}$, is added to the output of the final layer of the Transformer, after dimension reduction using a MLP, as follows:

$$\text{OmniNet}(X) = \text{Transformer}(X)_L + \text{MLP}(O_{\text{attention}}), \tag{4}$$

where $\text{Transformer}(X)_L$ represents the output of the Transformer model. The global attention in the OmniNet enables access to knowledge from the entire network, facilitating the capture of patterns that intertwine across different features. By reinforcing these correlations, the OmniNet improves its representational capacity and effectively captures the complex relationship between performance and preference. Furthermore, the global attention mechanism can be seen as a form of residual learning [97], which aids in gradient propagation and allows for end-to-end learning.

*4.4.3 Normalized Economy Difference based Optimization.* The win prediction task in the context of MOBA games can be framed as a binary classification problem. To optimize this task, we can utilize the standard binary cross-entropy (BCE) loss function:

$$L = -(\mathbf{y} \cdot \log(\hat{y}) + (1 - \mathbf{y}) \cdot \log(1 - \hat{y})). \tag{5}$$

Here, $\mathbf{y}$ represents the game outcome, and $\hat{y}$ denotes the model's prediction of the win rate. In MOBA games, the final game outcome is closely associated with the economy difference between teams [12]. A larger positive difference indicates a greater advantage for winning the game, while a negative difference suggests that the team is an underdog and more likely to lose the match [98]. By considering the economy difference as a key factor, the model can effectively predict the win rate and contribute to the overall fairness evaluation of potential assignment.

Based on the aforementioned observation, we enhance the win prediction task by incorporating information about the normalized economy difference (NED) into the loss function. The modified loss function is defined as follows:

$$L_{NED} = -(\alpha \cdot \mathbf{y} \cdot \log(\hat{y}) + (1 - \alpha) \cdot (1 - \mathbf{y}) \cdot \log(1 - \hat{y})). \tag{6}$$

Here, $\alpha$ represents the proportion of the team's economic difference, calculated as $\alpha = \frac{TE_1 - TE_2}{\max(TE_1, TE_2)}$, where $TE_1$ and $TE_2$ denote the overall economic values of the blue and red teams respectively. The purpose of incorporating $\alpha$ is to introduce dynamic label smoothing based on the specific battle context, improving the generalization ability of the proposed model. A positive value of $\alpha$ indicates that the blue team has an advantage, leading to a higher weight assigned to the blue team in the loss function. Conversely, a negative value of $\alpha$ suggests that the red team is favored, resulting in a higher weight assigned to the red team. By incorporating this economic information, we anticipate that the model will be better equipped to accurately predict battle outcomes and provide deeper insights into the dynamics of team performance in MOBA games.

## 5 OFFLINE EXPERIMENTS

In the offline experiment, our primary focus is to evaluate the performance of the win rate prediction model in assessing the ability of OwO to accurately predict game results, which serve as an indicator of fairness in player assignments. To this end, we aim to address the following research questions (RQs):

- **RQ1:** How does OwO perform compare to state-of-the-art methods in MOBA game win rate prediction?
- **RQ2:** How effective are each specific design of feature group and model component in OwO?
- **RQ3:** What impact does the NED loss make to the final win rate prediction?

We provide answers to these RQs in the following subsections.

## 5.1 Experiment Settings

*5.1.1 Baseline Methods.* In the offline experiments, we conducted a thorough evaluation of our proposed method by comparing it against 11 baseline models. Namely:

- **TrueSkill [19].** This probabilistic MMR model updates a player's rating solely based on the game outcome (win/lose). TrueSkill is widely utilized in online multiplayer games and serves as one of the most popular MMR algorithms.
- **TrueSkill2 [21].** TrueSkill2 builds upon its predecessor by incorporating additional features into the model, such as player experience and skill in other game modes. It typically converges faster and achieves higher accuracy compared to TrueSkill.
- **QuickSkill [11].** QuickSkill is a deep learning-based model for skill estimation that employs the MMR-Net to learn underlying patterns in player behavior and predict their skill scores in a given game. In this work, we utilized the TrueSkill2-based MMR-Net as the baseline method, as it exhibited superior performance in the original paper.
- **Logistic regression (LR) [99].** LR is a commonly used shallow model for binary prediction. Many previous works, such as [73, 100], have applied it to predict the final outcome in game battles.
- **EXtreme Gradient Boosting (XGboost) [72].** XGBoost is an implementation of gradient-boosted decision trees renowned for its speed, performance, and ability to handle large datasets. It has gained popularity in win prediction in games due to its high accuracy and support for parallel processing.
- **BalanceNet [61].** BalanceNet is a state-of-the-art network-based method used to predict which team will emerge victorious in online games. In line with the original paper, we employed an element-wise *tanh* activation function in the hidden vectors.
- **XGBoost with Stacking (XGBStack) [80].** XGBStack demonstrates that incorporating meticulous cross-features can provide additional relevant information to the classifier. As a baseline for comparison, we selected XGBStack, which is a meta-model that achieved the best performance in the original paper.
- **Multilayer Perceptron (MLP) [101].** MLP is a simple neural network architecture widely employed for various purposes. It consists of multiple layers of nodes, each connected to the next layer, and uses a feed-forward mechanism to propagate information through the network.
- **Long Short-Term Memory (LSTM) [102].** LSTM is a classical variant of recurrent neural network (RNN) that employs a gating mechanism to capture long-term sequence dependencies.
- **Transformer [90].** Transformers have gained prominence for their ability to model long-range dependencies in sequential data. Transformers utilize multi-head self-attention mechanisms to replace traditional RNN structures.
- **OmniNet [89].** OmniNet is a state-of-the-art variant of the transformer architecture that employs omnidirectional attention to connect all tokens across the entire network via self-attention. In this implementation of the proposed method, omnidirectional attention-based structures are utilized. It is also a subcomponent of the proposed OwO.

*5.1.2 Datasets.* Due to the limited availability of game features in publicly accessible datasets, we collected extensive industrial datasets from a popular MOBA game encompassing different game modes, including ranking and normal modes [3]. To ensure the reliability and accuracy of our analysis, we adopted a standard approach by randomly dividing the samples into a training set (80%) and a testing set (20%). Hyper-parameter tuning for our proposed model and baselines was performed using grid search on a small validation set, which accounted for approximately 10% of the training set. Summary statistics of the datasets are presented in Table 3. It is important to note that both normal and ranking game modes share the same game context and objectives. However, players tend to approach them differently. In the normal game mode, players may adopt a more casual approach, utilizing unskilled characters for practice and entertainment purposes. Conversely, players tend to exhibit a more serious attitude towards the ranking mode, as it provides a tangible grade to demonstrate their gaming prowess.

---

[3]In compliance with a non-disclosure agreement, we are unable to disclose the name of the game used in our study.

Table 3.  Statistics of the benchmark datasets.

| Dataset | #Training | #Validation | #Testing |
|---------|-----------|-------------|----------|
| Ranking | 30M       | 3M          | 8M       |
| Normal  | 4M        | 0.4M        | 1M       |

We would like to underscore that all data collection procedures adhered to relevant regulations and guidelines. Moreover, the dataset utilized in our study is devoid of any personally identifiable information pertaining to individual players. Consequently, the dataset has been subjected to thorough anonymization and de-identification processes, thus alleviating any potential privacy concerns associated with its utilization for research purposes. We prioritize data privacy and have diligently undertaken measures to ensure the ethical and responsible conduct of our research.

*5.1.3  Implementation Details and Evaluation Metrics.* The proposed model was implemented using TensorFlow [103], a widely used open-source machine learning platform. The offline experiments were conducted on a server equipped with 4 Tesla A100 GPUs. For the OmniNet component used in the TOE, SOE, and POE of our proposed method, we configured the hidden layer to have 3 layers, each with 256 units, and 10 attention heads. The final MLP in our model consisted of three hidden layers with hidden units of (256, 128). We experimented with learning rates in the range of 1e-3 to 1e-6 using the Adam optimizer [104].

Given that the win rate prediction problem can be framed as a binary classification task and the distribution of labels (*i.e.*, game outcomes) is balanced in both the normal and ranking datasets, we have opted to utilize the widely employed metric of *Accuracy* [79, 80, 105–107] for evaluating our proposed methods and baselines in the offline experiments. This metric is sufficient to provide an assessment of the models' ability to correctly predict the outcome of games.

## 5.2  Results Compared with Baselines (RQ1)

The accuracy of pre-match win rate prediction serves as a crucial indicator for assessing the fairness of performance in matchmaking. In Figure 6, we present the prediction accuracy of all models considered in this study on the test dataset. Observe that our OwO model consistently outperforms all other baseline models in both playing modes, affirming the effectiveness of its design. Specifically, OwO achieves a relative improvement of 2.57% to 18.45% (average 7.18%) on the ranking dataset and a relative improvement of 1.96% to 12.58% (average 6.24%) on the normal dataset. Specifically, the OwO model demonstrates a relative improvement ranging from 2.57% to 18.45% (with an average improvement of 7.18%) on the ranking dataset. Similarly, on the normal dataset, OwO achieves a relative improvement ranging from 1.96% to 12.58% (with an average improvement of 6.24%). The inclusion of dedicated components in OwO allows for the capture of cross-information from diverse features within large datasets. This approach surpasses the conventional feature engineering-based cross-feature design and enables the learning of more effective representations for outcome prediction. The higher prediction accuracy achieved by OwO translates to a higher precision in fairness assessment. These results highlight the potential of using OwO as a win rate prediction model to rank fairness effectively. By enhancing the matchmaking quality, OwO contributes to an improved player experience, ultimately benefiting the overall game ecosystem. Furthermore, the model-based approaches outperform the MMR-based methods (TrueSkill and TrueSkill2) in terms of accuracy. This finding supports the idea that relying solely on MMR for matchmaking is insufficient in accurately capturing the relationships between players, potentially leading to unfair match outcomes. Model-based methods, which incorporate additional dimensions of player features, demonstrate a clear advantage in this regard.

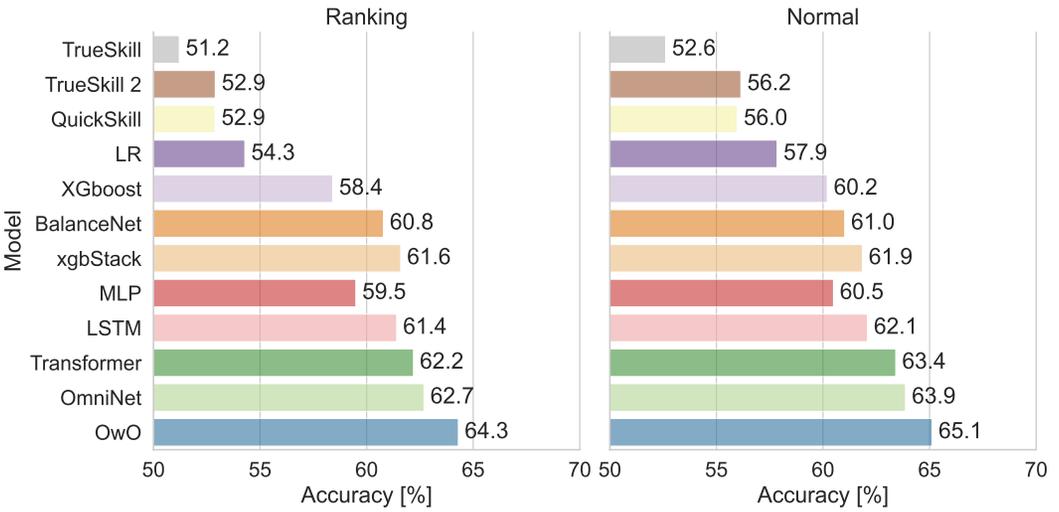

Fig. 6. The performance of pre-match win prediction methods on two benchmark datasets.

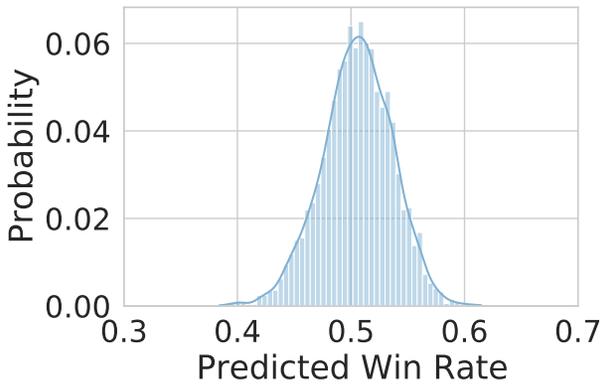

Fig. 7. Histogram of predicted win rate using the OwO model.

To assess the effectiveness of the proposed OwO in capturing subtle differences between different team sides, we conducted an experiment using manually generated samples. Specifically, we randomly selected 5,000 examples from the validation set and grouped all users on the blue team. For the opposing side, we constructed users with identical features to those of the blue team, resulting in equal inputs for both teams in each sample. We then generated a histogram of the predicted win rates, as depicted in Figure 7. The results indicate that the model's output is concentrated within a narrow range around 50%, with a mean of 50.5% (ranging from 44.5% to 56.6%, with a confidence interval of 95%). This demonstrates the stability of the proposed model and its ability to accurately predict game outcomes in balanced matches. Moreover, the mean predicted win rate for the blue team is 50.5%. This is not equal to the fair win rate of 50% but is closer to the actual win rate of the blue team online. This suggests that our model effectively captures the subtle differences between different team sides during training, leading to improved performance in the re-matchmaking system. These findings also provide valuable insights for game designers, as such variations in team sides are often influenced by game design factors such as the order of Ban/Pick and the positioning of neutral creatures.

## 5.3 Ablation Studies (RQ2)

Next, we perform a thorough ablation analysis to assess the impact of each feature group and model component design in CUPID. This analysis provides valuable insights into the contribution of these design choices towards improving the win rate prediction performance.

Table 4. Model structures comparison.

| Model groups | Ranking | Normal |
|---|---|---|
| OwO w/o TOE | 62.4% | 63.1% |
| OwO w/o SOE | 63.9% | 64.6% |
| OwO w/o POE | 62.9% | 63.8% |
| Complete OwO | 64.3% | 65.1% |

*5.3.1 Performance w.r.t. Model Structures.* To evaluate the individual effectiveness of the three OmniNet-based structures in our proposed OwO model, namely TOE, SOE, and POE, we conducted an ablation study using the ranking dataset. This study aimed to investigate how the accuracy performance changes when one of these components is removed. We replaced each OmniNet-based structure with a Multi-Layer Perceptron (MLP) to create modified versions of the OwO model. For example, OwO w/o TOE indicates the OwO model with TOE replaced by an MLP.

The results of the ablation study are presented in Table 4, which show the accuracy and relative improvement achieved by different models. Notably, the OwO model demonstrated the best performance, confirming the effectiveness of our purpose-built design. Specifically, when comparing OwO with and without TOE, OwO exhibited a significant relative improvement ranging from 3.0% to 3.2%. This highlights the power of omnidirectional attention-based structures in extracting high-level information from sequential data. In contrast, when comparing OwO without SOE, the relative improvements were smaller, ranging from 0.6% to 0.8%. This can be attributed to the fact that the input spatial features were processed through feature engineering and transformed into one-hot representations, making it easier for even a simple MLP model to learn good representations. This finding aligns with previous research findings. Furthermore, even in comparison to OwO without POE, OwO still achieved a decent relative improvement ranging from 2.0% to 2.2%. This suggests that OmniNet can effectively learn cross-domain knowledge between different users' embeddings through globally-perceptive attention.

*5.3.2 Feature groups comparison.* The experimental results using different combinations of feature groups on the ranking dataset are presented in Table 5. It is evident that the complete OwO model achieved significantly better performance compared to the results obtained with incomplete feature sets. This observation indicates that our proposed model is effective in integrating diverse feature data, extracting advanced representations, and improving prediction accuracy.

Table 5. Results of proposed OwO *w.r.t* different feature groups.

| Featrue Groups | Ranking | Normal |
|---|---|---|
| Team Featue | 56.3% | 58.1% |
| Short-term Featues | 58.9% | 60.8% |
| Long-term Featues | 58.2% | 60.6% |
| Team Featues + Short-term Featues | 60.3% | 63.0% |
| Team Featues + Long-term Featues | 60.1% | 61.9% |
| Short-term Featues + Long-term Featues | 60.6% | 62.3% |
| All Features | 64.3% | 65.1% |

Table 6. The accuracy comparison of MMR-based and OwO-based prediction on game outcomes.

|  | MMR | OwO |
|---|---|---|
| One-Step Matchmaking | 52.83% | 63.19% |
| Two-Step Matchmaking | 48.99% | 52.55% |

Furthermore, we noticed that models utilizing short-term features outperformed those using long-term features. This suggests that recent features of players play a more crucial role in predicting game outcomes. The inclusion of recent features allows for the inference of long-term player characteristics, such as location preferences, which can be effectively captured by our model. Interestingly, models only incorporating team features, which provide a higher-level and more abstract representation of players, exhibited the lowest performance. This finding emphasizes the importance of profiling players from additional perspectives, as done in our proposed Cupid framework. By considering a more comprehensive range of player attributes, we can significantly enhance win prediction accuracy and improve the fairness assessment in the matchmaking process.

*5.3.3 One-step vs. two-step matchmaking.* To further demonstrate the superiority of the two-step matchmaking approach, we conducted an offline analysis of the prediction accuracy for both MMR-based and OwO-based methods using the online dataset. The results, presented in Table 6, reveal that in one-step matchmaking, OwO significantly outperforms MMR-based methods that use TrueSkill, indicating a substantial improvement in match fairness. In two-step matchmaking, the MMR-based method is nearly ineffective, with an accuracy below 50% (equivalent to random guessing accuracy). Furthermore, the two-step matchmaking approach reduces the overall predictability of match outcomes, thereby enhancing match fairness.

## 5.4    Performance Comparison *w.r.t.* NED loss (RQ3)

We now shift our focus to evaluating the impact of the NED loss on prediction performance. Table 7 presents the prediction accuracy of OwO trained with the BCE loss and OwO trained with the NED loss (referred to as $OwO_{NED}$). It can be observed that $OwO_{NED}$ achieves improved performance compared to OwO, indicating that incorporating a loss function based on economic differences can lead to more effective results, as this indicator is positively correlated with game outcomes.

To provide a more comprehensive analysis of the NED loss, Figure 8 presents three subplots. Subplot 8(a) depicts the evolution of the loss during the training process, while subplot 8(b) shows the accuracy on the validation dataset throughout training. It can be observed that both OwO and $OwO_{NED}$ exhibit a decreasing trend in loss as the number of epochs increases, although OwO demonstrates a more significant reduction. However, the accuracy of OwO initially increases but then quickly declines, indicating the presence of overfitting. In contrast, the accuracy of OwO$_{NED}$ remains relatively stable after reaching its peak and does not show a significant decrease during training. This observation suggests that incorporating Eq. (6) in training the model can effectively regularize it, leading to improved generalization and overall performance.

Table 7. The performance comparison of OwO and OwO$_{NED}$.

| Model | Ranking | Normal |
|---|---|---|
| OwO | 64.3% | 65.1% |
| OwO$_{NED}$ | 64.5% | 65.3% |

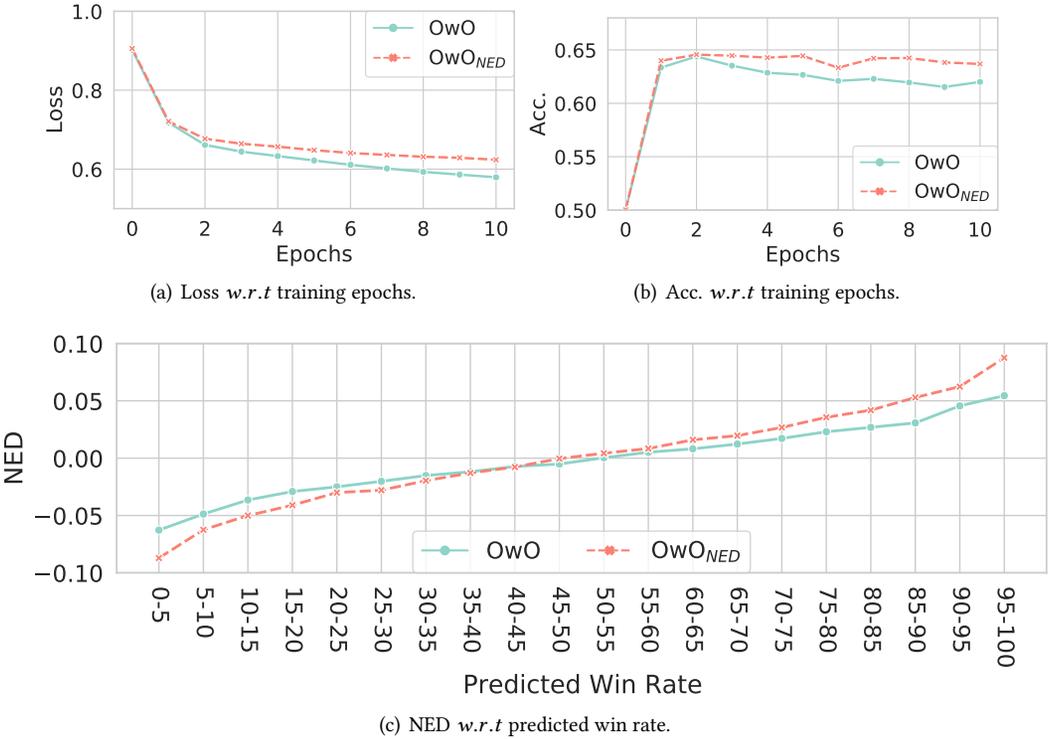

(a) Loss *w.r.t* training epochs.

(b) Acc. *w.r.t* training epochs.

(c) NED *w.r.t* predicted win rate.

Fig. 8. Comparisons of OwO with OwO$_{NED}$ from difference perspectives.

Furthermore, subplot 8(c) illustrates the relationship between economic difference and the predicted win rate. It can be observed that after incorporating the NED loss, the correlation between the model's predicted win rate and the economic difference between the two teams becomes stronger. This has two implications. Firstly, it enables a more distinct differentiation between fair and unfair games based on the predicted win rate, thereby facilitating the ranking process. Secondly, it tightens the fairness with respect to team economic, reducing the occurrence of overwhelming games in terms of economic difference when the NED loss is employed.

## 6 ONLINE EXPERIMENTS

We implemented the Cupid framework as a re-matchmaking system within a well-received mobile MOBA game, catering to a substantial player base numbering in the millions. To gauge its performance, we executed A/B testing within a live production environment and conducted assessments utilizing comprehensive metrics that encapsulate accessibility, usability, and engagement perspectives [108, 109], which are fundamental in the HCI domain. In addition, we also evaluate whether CUPID can achieve high efficiency for the re-matchmaking, ensuring scalability and negligible additional processing time that would impact players' experience. In particular, the online experiments were designed to address the following research questions:

- **Usability (RQ4):** Does CUPID exhibit commendable usability, ensuring fair matchmaking and user satisfaction with their assigned positions?
- **Engagement (RQ5):** Can the proposed CUPID system enhance the overall gaming experience and engagement for players, leading to improved player behaviors?
- **Accessibility (RQ6):** Can the proposed CUPID system be effectively cater to different user groups, including those with varying skill levels and from different regions?
- **Efficiency (RQ7):** Can the proposed CUPID system operate efficiently online, ensuring that players' experience during the matchmaking period is not impacted?

We utilize Normalized Absolute Economy Difference (NAED), Crushing Rate (CR), and Misassignment Ratio (MR) to evaluate the system's Usability (RQ4). For assessing CUPID's Engagement (RQ5), we employ Average Away From Keyboard (AFK) level, Active Quit Rate (QR), Active Quit Time (QT), Bounce Rate (BR), and Average Number of Battles (ANOB). In gauging its Accessibility (RQ6), we consider whether the positive effect of CUPID can be accessible to players with different skill levels and position preference. Furthermore, the Efficiency (RQ7) of CUPID is evaluated via Success Rate (SR), System Responsive time (SRT), Query per Second (QPS) and Efficiency Ratio (ER). Detailed information is presented in the subsequent subsections.

## 6.1 Deployment Scale and Experiment Method

To evaluate the effectiveness of our proposed system in a real-world MOBA game matchmaking environment, we conducted a rigorous A/B testing within the ranking mode of a popular game with a substantial player base of tens of millions. The testing was carried out over a period of 30 days, during which we randomly assigned 10% of the matches to the experiment group, where the proposed re-matchmaking system was implemented. The remaining matches served as the control group and followed the traditional MMR-based matchmaking approach. The ranking mode was specifically chosen for the evaluation due to the heightened interest among players in being able to select their desired positions and ensuring game balance within this mode. All subsequent online experiments compare the two-step re-matchmaking driven by CUPID with the traditional one-step MMR-based matchmaking system using TrueSkill.

## 6.2 Usability (RQ4)

Usability in the context of matchmaking in MOBA games pertains to the effectiveness of the matchmaking system in realizing fair, competitive, and enjoyable games for both teams of players, while concurrently ensuring players can assume their desired position roles. To assess the usability of CUPID, we employ a set of metrics focusing on game fairness and position satisfaction, namely:

- **Normalized Absolute Economy Difference (NAED)**: This metric reflects the state of the game by measuring the difference in team economy. A larger NAED indicates a more imbalanced game. We observed the NAED at the 5th minute (NAED@5) and 15th minute (NAED@15) of the battle to assess the balance during the early and late stages, respectively.
- **Crushing Rate (CR)**: In MOBA games, when the economy difference reaches a certain threshold, it becomes difficult to reverse the game, resulting in a "crushing" scenario. The crushing rate is an important indicator for assessing game balance. Similar to the NAED

Table 8. Relative improvement results on usability metrics.

| Metics | Relative Improvement | Metics | Relative Improvement |
|--------|---------------------|--------|---------------------|
| NAED@5 | 2.18% | NAED@15 | 3.31% |
| CR@5 | 10.54% | CR@15 | 8.23% |
| MR@1 | 4.13% | MR | 25.27% |

metric, we measured the crushing rate at the 5th minute (CR@5) and 15th minute (CR@15) of the game.

- **Misassignment Ratio (MR)**: Prior to the game, players can choose their preferred positions for the upcoming match. The misassignment ratio measures the proportion of players whose actual assigned position does not match their pre-selected position. This metric directly reflects the satisfaction of players with the position assignment by the matchmaking system, particularly in ranked mode. We selected the proportion of players who did not get their pre-selec position (MR@1) and the average MR as the most stringent indicator to assess players' position satisfaction.

The first two metrics serve as important indicators for game fairness, and the last MR is employed to assess the players' position satisfaction. They are widely employed for game designers. Note that the lower these metrics are, the better of the performance.

The results of the online experiment are presented in Table 8, with the actual numbers of the metrics withheld due to confidentiality requirements. Our proposed methods demonstrate significant improvements compared to the legacy matchmaking system across all metrics. Specifically, the experiment group, which utilized the re-matchmaking system Cupid, achieved a relative improvement of 2.18% and 10.54% in NAED@5 and CR@5, respectively. Additionally, there was a relative improvement of 3.31% and 8.23% in NAED@15 and CR@15, respectively. These results indicate that Cupid can effectively enhance game fairness, resulting in a smoother game experience during both the early and late stages. The smaller economy differences imply a reduced likelihood of a "crushing" scenario, highlighting the improved balance in team performance facilitated by Cupid. Ultimately, this leads to increased player satisfaction and enjoyment in MOBA battles.

Furthermore, the experiment group exhibited significant improvements in terms of position satisfaction. The utilization of Cupid resulted in a relative improvement of 4.13% in MR@1 and an overall improvement of 25.27% in MR. These findings validate the effectiveness of the pre-filtering phase in Cupid, where *(i)* players are more likely to assigned to the position they pre-select, and *(ii)* overall they are assigned their preferred position. In essence, Cupid achieves dual benefits, enhancing both game fairness and players' position satisfaction, ultimately contributing to an enhanced gaming experience. This demonstrates the superior usability of Cupid when compared to the one-step matchmaking.

## 6.3 Engagement (RQ5)

In the context of MOBA games, engagement metrics gauge the matchmaking system's ability to capture and sustain players' attention, interest and retention, as well as its impact on players' in-game behaviors. To scrutinize this facet of Cupid, we choose various engagement metrics to elucidate user feedback, including:

- **Average Away From Keyboard (AFK) level**: AFK level signifies the duration players spend away from the keyboard in the game, with higher levels indicating longer durations. Players typically go AFK in response to unpleasant experiences. Thus, a lower average AFK level indicates an improved gaming experience.

Table 9. Relative improvement results on different engagement metrics.

|  | AFK level | QR | QT | BR | ANOB |
|---|---|---|---|---|---|
| Relative Improvement | 0.68% | 9.77% | 2.14% | 7.42% | 82.79% |

- **Active Quit Rate (QR)**: The active quit rate represents the ratio of players who initiate matchmaking but actively cancel the process. Active quitting is commonly driven by an inability to tolerate waiting times, making this metric indicative of negative player feedback.
- **Active Quit Time (QT)**: Active quit time is the duration between a player initiating matchmaking and actively canceling it. Players willing to tolerate longer waiting times often believe it leads to better match quality, making this metric a positive feedback indicator.
- **Bounce Rate (BR)**: Bounce rate is the ratio of players who continue playing after completing a game in a short period. A good gaming experience in a match usually prompts players to start another game, making this metric a positive feedback indicator.
- **Average Number of Battles (ANOB)**: This metric represents the total number of battles a player engages in within a specific period, serving as a direct measure of player activity level.

Among these metrics, AFK, QR, QT and BR emphasize players' short-term feedback for individual matches or the matchmaking process, while ANOB can be viewed as reflecting players' long-term feedback on the game ecosystem. All of them serve as vital metrics in evaluating player engagement.

We present the relative improvement compared to the one-step matchmaking for all engagement metrics in Table 9. It is observed that Cupid significantly enhances metrics in this context, demonstrating its considerable ability to improve players' engagement. Specifically, the average AFK level in the experimental group decreased by 0.68%, indicating a reduction in toxic behaviors and a better playing environment provided by Cupid. The QR witnessed a substantial relative decrease of 9.77%, and QT dropped by 2.14%, implying that players are more satisfied with the positions they are assigned, making them less likely to quit the game.

Interestingly, we observe that the BR increased by 7.42% with the use of Cupid. This suggests that more players continued playing after completing a game with re-matchmaking, indicating a more appealing and joyful gaming experience that encourages them to play again. Importantly, players who used Cupid demonstrated a remarkable relative improvement of 82.79% in the ANOB. This indicates a higher level of player engagement, suggesting that the use of Cupid led to increased participation.

In summary, the results of the online A/B experiment suggest that the implemented changes have significantly improved player engagement. The decrease in AFK level, QR, and BR, along with the increase in QT and ANOB, all indicate a more positive gaming experience for players. These improvements are attributed to an optimized re-matchmaking by Cupid, providing more balanced matches and an overall better gaming experience.

## 6.4 Accessibility (RQ6)

Accessibility in matchmaking within the context of MOBA games refers to the efforts made by game developers to create a fair and enjoyable gaming experience for *all* groups of players, taking into consideration factors such as skill level, position preference, and other relevant metrics. In consideration of accessibility for individuals with disabilities or special needs, we have chosen not to delve into this aspect, as it is less pertinent in the context of matchmaking. Cupid is designed to accommodate all player groups, ensuring high accessibility across various players demographics.

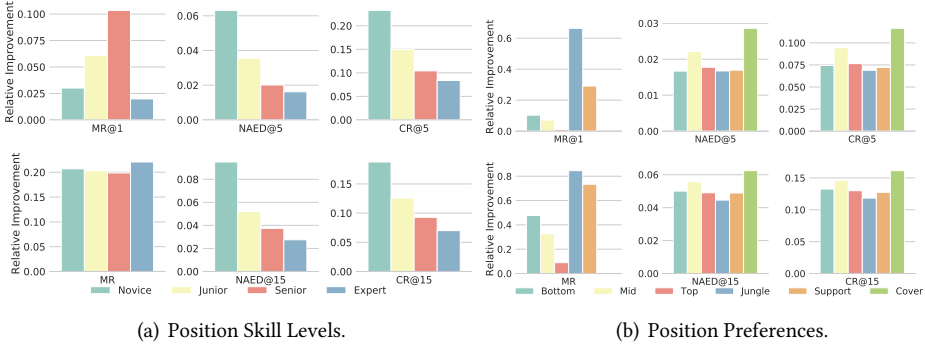

(a) Position Skill Levels.

(b) Position Preferences.

Fig. 9. Relative improvement on A/B testing for players with different user groups.

In this analysis, we meticulously examine the NAED, CR, and MR metrics across diverse player profiles and needs, aiming to evaluate the effectiveness and accessibility of Cupid across different user groups. These groups include:

- **Players with different skill levels**: We categorize players into four skill levels based on their ranking: novice, junior, senior, and expert.
- **Players with different position preferences**: Players are grouped based on their pre-selected positions.

The breakdown results of NAED, CR, and MR of different players skill levels are presented in Figure 9(a). Interestingly, we observed a consistent trend where the improvement in fairness-related metrics becomes more pronounced as the player level decreases. This can be attributed to the fact that higher-level players are typically more versatile and capable of performing well in multiple positions. Conversely, the results suggest that Cupid is particularly beneficial for low-level players, potentially enhancing their retention rate in the early stages of the game. Furthermore, while our proposed framework demonstrates relative improvement in satisfaction-related metrics across all player groups, the gains are relatively smaller in the expert player group. This is because in high-level games, players often gravitate towards positions that have a greater impact on game results, such as Mid and Bottom. As a result, the preferred position distribution among high-level players becomes more concentrated, making position allocation more challenging. However, it is worth noting that the MR metric exhibited significant improvement across all player groups, with an average improvement of approximately 20%. This reaffirms our previous findings and highlights the effectiveness of the proposed model in position assignment. Overall, the results indicate that players of all skill levels experience the positive effects introduced by Cupid, showcasing its high accessibility.

We further analyze the results based on players' position preferences, as illustrated in Figure 9(b). The colors in the figure correspond to different position preferences among players. The findings reveal notable enhancements in our proposed framework across all metrics for players with diverse position preferences. Particularly, the NAED and CR metrics exhibit relatively consistent improvements across all player groups, while the MR metric displays more variability. In more detail, the MR metric demonstrates significant advancements for players favoring the Jungle and Support positions, whereas the improvements are comparatively lower for players who selected the Bottom, Mid, and Top positions. This discrepancy can be attributed to the higher number of players already preferring the Bottom, Mid, and Top positions in our game, due to their high popularity, resulting in increased conflicts and posing challenges for achieving substantial improvements in these positions. Nevertheless, Cupid demonstrates improvement for players with different position preferences, thereby catering to players with diverse needs.

Table 10. Statistics of system latency of Cupid in real production deployment.

| P10 | P50 | P90 | P99 | Mean |
|---|---|---|---|---|
| 65.58 ms | 102.11 ms | 315.02 ms | 685.50 ms | 155.41 ms |

Despite the metrics selected extending beyond traditional accessibility considerations, focusing on groups such as those with disabilities, Cupid showcases broad coverage across various player groups. This underscores its high accessibility from a different perspective.

## 6.5 Efficiency (RQ7)

We also evaluate the efficiency performance of the overall system to ensure that Cupid can meet the real-world system's requirement for online responsiveness. This evaluation is crucial as extended matchmaking times can lead to player impatience and potential abandonment of the queue. To this end, we select the following metrics:

- **Success Rate (SR):** The success rate of a matchmaking system refers to the ratio of correctly processed player requests by the system, encompassing both the initiation and termination of match requests.
- **System Responsive Time (SRT):** This evaluates the total latency of the matchmaking process.
- **Query per Second (QPS):** This evaluates how many requests Cupid can handle per second.
- **Efficiency Ratio (ER):** This is defined as the ratio of the number of assignments that meet the satisfaction threshold and pass the pre-filtering process to the total number of full assignments.

In the online A/B experiment, the Cupid system demonstrated a relative improvement of 0.05% in the SR metric. These results signify that Cupid exhibits improved accessibility, facilitating more efficient matchmaking for players. Notably, both Cupid and the original matchmaking system attained absolute SR values surpassing 99.99%, underscoring the excellent accessibility of our matchmaking system. It consistently and accurately processes user requests in the overwhelming majority of instances.

Table 10 presents the statistics of the SRT of Cupid. On average, Cupid demonstrates a low responsive time of approximately 150 ms. Even at the P99 percentile, the responsive time remains below 700 ms. These results indicate that Cupid performs efficiently and is capable of responding to matchmaking requests in a timely manner, ensuring a smooth and responsive user experience. Players are unlikely to perceive any noticeable delay caused by Cupid, especially considering that it typically takes more than 1 second for a player to click the "Accept" button (as shown in Figure 3(b)). This duration is significantly longer than the overhead of Cupid. Therefore, we can confidently conclude that our re-matchmaking framework, Cupid, does not prolong the matchmaking time and can be seamlessly integrated into the existing matchmaking pipeline.

Additionally, we conducted stress tests specifically targeting the model service module, which is the component with the highest computational demands in our proposed system. These tests were carried out on a dedicated server equipped with 4 CPU cores. During the stress testing phase, Cupid flawlessly processed all incoming requests, demonstrating a 100% success rate. The purpose of these tests was to assess the system's ability to handle high volumes of requests and respond within acceptable latency limits. Table 11 presents the statistical information obtained from the stress tests, including the queries per second (QPS) and system latency, for different numbers of assignments (#assignments) and concurrency levels (#concurrency). The metric #assignments refers to the number of assignments included in a single request, indicating the number of assignment predictions made by the model in one batch.

Table 11. The QPS and statistics of latency results in the stress tests with different #concurrency and #assignments levels. All time measurements for time overhead are in milliseconds.

| #Assignment | #Concurrency | P10 | P50 | P90 | P99 | Mean | QPS |
|---|---|---|---|---|---|---|---|
| 1 | 20 | 83.89 | 136.38 | 243.59 | 376.74 | 151.03 | 133.81 |
| 4 | 20 | 127.68 | 216.64 | 371.11 | 614.71 | 231.59 | 84.56 |
| | 50 | 241.60 | 433.87 | 789.86 | 973.02 | 469.76 | 84.18 |
| | 100 | 70.16 | 446.14 | 966.24 | 994.54 | 507.47 | 82.03 |
| 8 | 20 | 172.54 | 289.30 | 493.03 | 791.31 | 310.43 | 63.90 |
| | 50 | 201.24 | 492.96 | 833.86 | 985.12 | 517.15 | 61.94 |
| | 100 | 168.79 | 389.12 | 803.02 | 975.01 | 436.74 | 61.15 |
| 16 | 20 | 212.54 | 360.34 | 600.59 | 900.33 | 381.11 | 52.47 |
| | 50 | 160.50 | 574.91 | 882.48 | 990.22 | 552.18 | 51.76 |
| | 100 | 163.68 | 570.96 | 879.03 | 988.06 | 552.19 | 50.83 |

On the other hand, the metric #concurrency represents the number of requests sent to the model server simultaneously, indicating the level of parallelism in processing multiple assignment predictions. These metrics are crucial for evaluating the efficiency and scalability of the model server in handling a high volume of matchmaking requests. The results demonstrate that the performance of the model service module is in line with our expectations and meets the requirements of the system. Even under extreme scenarios with a concurrency level of 100, the system is capable of efficiently handling a large number of requests while maintaining low latency. It is worth noting that the system's efficiency is further enhanced by its ability to predict win rates for multiple assignments simultaneously within a single request. This capability, which closely resembles real-world scenarios where multiple players are involved, benefits from the optimization achieved through batched computation. By processing multiple assignments in parallel, the system can maximize resource utilization and improve overall throughput.

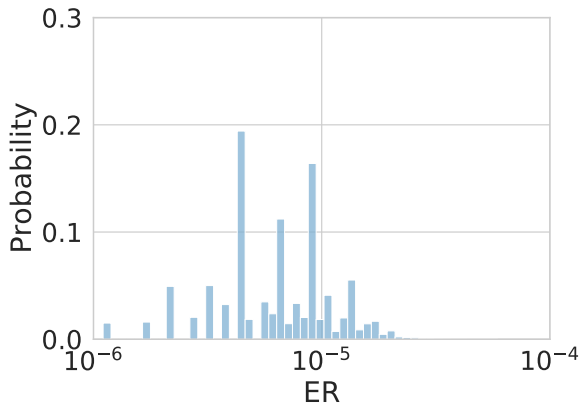

Fig. 10. Histogram of Efficiency Ratio (ER) in online data.

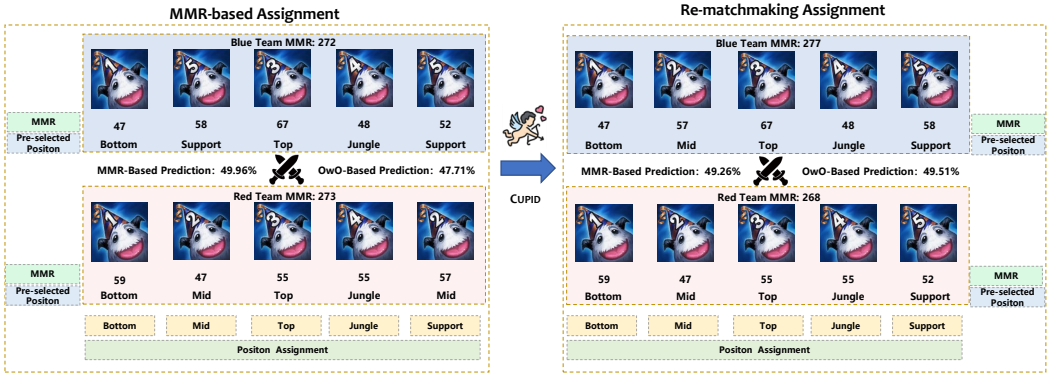

Fig. 11. A case Study of the effect of the re-matchmaking process of Cupid.

In order to assess the efficiency impact of the pre-filtering process, we introduced the Efficiency Rate (ER) metric, *i.e.*, $ER = \frac{N_{act}}{N_{full}}$ defined as the ratio of the number of assignments that meet the satisfaction threshold and pass the pre-filtering process (denoted as $N_{act}$) to the total number of full assignments (denoted as $N_{full}$). A lower ER indicates a more effective pre-filtering process, as it means fewer assignments need to be predicted by the OwO model, resulting in higher system efficiency. Figure 10 illustrates the histogram of the ER metric, which is derived from actual online data obtained from the experiment group in the A/B testing. The x-axis is presented on a logarithmic scale. As depicted in Figure 10, our strategy significantly reduces the computational complexity to a minimal range, ranging from $10^{-6}$ to $10^{-4}$ (with an average of $7 \times 10^{-6}$). This implies that without the pre-filtering process, the number of requests to the model would increase by approximately $10^6$, which is not feasible for online deployment. These results further validate the effectiveness of the pre-filtering process in Cupid, as it simultaneously improves position satisfaction and system efficiency.

In conclusion, the integration of a satisfaction-based selection strategy into the online re-matchmaking framework has resulted in significant efficiency improvements, with the system capable of handling a high volume of requests in the order of millions. The re-matchmaking process facilitated by Cupid exhibits negligible perceivability among players, safeguarding their matchmaking patience and overall gaming experience.

## 7 DISCUSSION

In the discussion section, we elucidate how the two-stage matchmaking, implemented in Cupid, outperforms its one-stage counterpart through a detailed case study. Additionally, we explore the generalizability of the proposed framework, delve into its limitations, distill lessons learned, and outline strategies for enhancing Cupid in future iterative designs.

### 7.1 Case Study

In Figure 11, we present a case study conducted in a real online game environment to illustrate how Cupid leverages the re-matchmaking process to account for player heterogeneity, achieving fair and satisfying competition and outperforming the traditional MMR-based one-step matchmaking approach.

Before the competition, the initial preferred positions for players in the blue and red teams were [[Bottom, Support, Top, Jungle, Support], [Bottom, Mid, Top, Jungle, Mid]], with corresponding MMRs of [[47, 58, 67, 48, 52], [59, 47, 55, 55, 57]]. However, in the MMR-based assignment (step 1), some players were assigned to different roles from their preferred positions. This mismatch is not considered by the MMR, resulting in a predicted fair win rate, as the two teams have close total MMR (272 vs 273). In contrast, the OwO model takes this deviation into account and identifies that this may not be a fair game (47.71% win rate).

After the reassignment of positions by the Cupid system, the red and blue teams exchange one player, aligning everyone with their preferred positions. Now, both MMR and OwO predict this as a fair game, with the predicted win rates fairly close to 50%. In this match, we observed it as a fairly competitive game in our post-analysis. This real-world case distinctly illustrates that the re-matchmaking process significantly enhances game fairness by incorporating players' position preferences. This superiority positions it as a more effective solution compared to the one-step MMR-based approach that determines team assignments. It is important to note that Cupid considers much richer features than just preferred positions for team and position assignments, as discussed in Section 4.2, comprehensively addressing players' heterogeneity to enhance matchmaking quality and improve the overall player experience.

## 7.2 Generalizability

While the primary design of Cupid is for 5v5 MOBA matchmaking games, such as LoL, Dota, and Mobile Legends, its adaptability extends beyond this specific context. The framework can readily be customized for other MOBA game modes and, more broadly, for diverse game types that necessitate team or position assignments to achieve equitable matchmaking. Only minor adjustments suffice to adapt Cupid to various games beyond its primary design for 5v5 MOBA matchmaking. Specifically,

- **Other 5v5 Matches**: For games akin to MOBA, the fundamental model structure remains unaltered, necessitating only the incorporation of game-specific features into the model's input.
- **PvP with 2 Teams**: Encompassing scenarios with balanced or unbalanced team sizes, such as FreeStyle Street Basketball (3v3) [110] and Evolve (1v4) [111], this category demands adjustments not only to input features but also to the model's structure to accommodate varying numbers of players. For example, in Evolve, where teams vary in size, features require restructuring to represent individual players while retaining information on distinct player classes.
- **Multi-Team Battles**: Examples encompass games like PlayerUnknown's Battlegrounds (PUBG) [112], where numerous players are divided into 16 teams. Adapting the Cupid model to such battles necessitates modifications to features, input structure, and output format. Here, the output transformation involves transitioning from a single digit to multiple probability digits (*e.g.*, 16), utilizing softmax activation. Each neuron's output then reflects the probability of a team winning.

Note that the satisfaction filter may also be tailored to specific games. Overall, Cupid is adaptable to other games that require position assignment with minor updates. This highlights the remarkable generalizability of the Cupid framework.

## 7.3 Limitations & Lesson Learned

To maintain a concise and lucid framework, our model's input is specifically designed to handle structured features, omitting modules for processing unstructured inputs such as graph data. Nevertheless, crucial features like player relations networks and the synergistic counter relationships among champions are inherently graph-based, and are presently overlooked in the existing framework. In future iterations, we aim to integrate network embedding techniques [17, 113, 114] to harness the wealth of network information available in games. This enhancement is intended to elevate the prediction model's performance and further refine the capabilities of Cupid.

Moreover, the OwO model currently predicts a single objective, specifically the win probability of one team. This singular focus makes OwO potentially sensitive to noise or outliers and limits its optimization to fairness. In future research, we plan to explore the implementation of multi-objective learning techniques [115, 116] to concurrently optimize satisfaction, game fairness, and matchmaking efficiency. This approach aims to provide a multi-faceted optimization framework for comprehensive objectives in matchmaking quality. These refinements are anticipated to fortify our framework, rendering it more robust, efficient, and adept in addressing the diverse challenges of matchmaking in MOBA games.

Finally, we acknowledge the significance of further evaluating the accessibility of Cupid for specific user groups. To this end, we have designed a questionnaire, presented in Appendix Section A, to assess how accessible Cupid is to different player groups, taking into account factors such as skill levels, position preferences, *etc.*. We are currently developing a questionnaire delivery tool for this purpose and plan to share the evaluation results in our future work.

### 7.4 Iterative Design

Although Cupid has demonstrated remarkable performance in both offline and online evaluations, continuous improvement is essential. We intend to enhance its capabilities through various measures, including the utilization of Testing Servers, conducting A/B Testing, and executing Questionnaire Surveys. These steps aim to fortify Cupid as a more robust, user-centric, and refined matchmaking system. Specifically:

- **Testing Servers:** We will establish a dedicated testing environment where player actions won't impact the official game environment. This will be used to assess new improvements on Cupid with a subset of players, allowing for the collection of feedback and opinions for system enhancement [117], as well as the analysis of system bugs and faults.
- **Continuous A/B Testing:** We will consistently deploy the new system to a subset of players within the official game environment using randomized A/B testing. By comparing data, such as matchmaking time and skill balance, between the groups using the new and current systems, we can evaluate the effectiveness of the new system. This approach helps us determine the superiority of the new system and facilitates a gradual introduction to minimize the impact on players.
- **Questionnaire Survey:** We will design a questionnaire survey to elicit players' opinions and suggestions regarding the matchmaking system. The questionnaire will be distributed through in-game announcements following matches and will include inquiries about matchmaking time, match quality, and other relevant aspects. Analyzing the survey results will provide insights into players' needs and expectations, guiding corresponding improvements based on their feedback. A questionnaire demo is provided in the Appendix Section A in the supplementary material.

By employing these three methods in tandem, we can gather comprehensive data and feedback, facilitating the ongoing enhancement of the game matchmaking system and the overall improvement of players' gaming experiences.

## 8 CONCLUSION

MOBA games have amassed a considerable player base and have drawn significant attention within the HCI research community. The quality of the matchmaking system in MOBA games is pivotal in ensuring a positive player experience. In this study, we introduce a novel re-matchmaking framework, referred to as Cupid, designed to address the shortcomings of traditional matchmaking systems that solely depend on unilateral MMR for team and player assignment. The primary aim of Cupid is to optimize team and position assignments to enhance the overall matchmaking quality. To accomplish this, Cupid integrates two key components. Initially, a pre-filtering process predicated on position satisfaction is utilized to guarantee that players are assigned to their preferred positions whenever feasible. This approach augments player satisfaction by taking into account their individual preferences. Subsequently, Cupid incorporates a custom win rate prediction model OwO, which functions as a fairness evaluator and ranker. OwO predicts the win rate for potential assignments, facilitating a more balanced match and improved fairness.

The efficacy of Cupid was evaluated through a combination of offline assessments and online A/B testing within prominent online MOBA games boasting millions of players. The results demonstrated noteworthy improvements in position satisfaction and game fairness, as gauged by critical HCI metrics encompassing usability, accessibility, and engagement. Our observations indicate that matches characterized by greater fairness contribute to elevated user satisfaction, thereby fostering a heightened inclination to partake in successive matches. To the best of our knowledge, Cupid stands as the pioneering re-matchmaking framework tailored for large-scale MOBA games.

## 9 ETHICS STATEMENT

In this paper, we present a deep learning-based re-matchmaking system, which includes an offline training process using collected features. We prioritize data privacy, as all features collected in this study do not contain personal information of players. In summary, the proposed re-matchmaking system is designed to connect individuals based on compatible traits, interests, and preferences. We are committed to ensuring that our technology upholds ethical standards and respects the privacy, autonomy, and diversity of our users.

## ACKNOWLEDGMENTS

This work was supported in part by National Natural Science Foundation of China under Grant No. 62102265, in part by Stable Support Project of Shenzhen under Grant No. 20231120161634002, in part by Natural Science Foundation of Guangdong Province of China under Grant No. 2022A1515011474.

## A  A PROVEN QUESTIONNAIRE FOR ASSESSING CUPID'S ACCESSIBILITY

| Sl. | Questions | Agreement |
|-----|-----------|-----------|
| Please rate your agreement with the following statements on a scale of -5 to 5. 5 indicates strong agreement and -5 indicates strong disagreement. | | |
| 1.1 | How would you rate the overall accessibility of the game matchmaking system? | -5 ☐ -4 ☐ -3 ☐ -2 ☐ -1 ☐ 0 ☐ 1 ☐ 2 ☐ 3 ☐ 4 ☐ 5 ☐ |
| 1.2 | How easy was it to navigate through the menus and options in the game matchmaking system? | -5 ☐ -4 ☐ -3 ☐ -2 ☐ -1 ☐ 0 ☐ 1 ☐ 2 ☐ 3 ☐ 4 ☐ 5 ☐ |
| 1.3 | Did you find the instructions and guidance provided in the game matchmaking system clear and understandable? | -5 ☐ -4 ☐ -3 ☐ -2 ☐ -1 ☐ 0 ☐ 1 ☐ 2 ☐ 3 ☐ 4 ☐ 5 ☐ |
| 1.4 | Are you satisfied with the position you were assigned? | -5 ☐ -4 ☐ -3 ☐ -2 ☐ -1 ☐ 0 ☐ 1 ☐ 2 ☐ 3 ☐ 4 ☐ 5 ☐ |
| 1.5 | Are you satisfied with the position you chose? | -5 ☐ -4 ☐ -3 ☐ -2 ☐ -1 ☐ 0 ☐ 1 ☐ 2 ☐ 3 ☐ 4 ☐ 5 ☐ |
| 1.6 | Are you satisfied with the hero you chose? | -5 ☐ -4 ☐ -3 ☐ -2 ☐ -1 ☐ 0 ☐ 1 ☐ 2 ☐ 3 ☐ 4 ☐ 5 ☐ |
| 1.7 | Are you satisfied with your teammates' performance? | -5 ☐ -4 ☐ -3 ☐ -2 ☐ -1 ☐ 0 ☐ 1 ☐ 2 ☐ 3 ☐ 4 ☐ 5 ☐ |
| 1.8 | Compared to your level, do you think the average level of your teammates is high? | -5 ☐ -4 ☐ -3 ☐ -2 ☐ -1 ☐ 0 ☐ 1 ☐ 2 ☐ 3 ☐ 4 ☐ 5 ☐ |
| 1.9 | Compared to your level, do you think the average level of your opponents is high? | -5 ☐ -4 ☐ -3 ☐ -2 ☐ -1 ☐ 0 ☐ 1 ☐ 2 ☐ 3 ☐ 4 ☐ 5 ☐ |
| 1.10 | Do you think the game duration was too long? | -5 ☐ -4 ☐ -3 ☐ -2 ☐ -1 ☐ 0 ☐ 1 ☐ 2 ☐ 3 ☐ 4 ☐ 5 ☐ |
| 1.11 | Do you think the result of the game was destined in the early stage? | -5 ☐ -4 ☐ -3 ☐ -2 ☐ -1 ☐ 0 ☐ 1 ☐ 2 ☐ 3 ☐ 4 ☐ 5 ☐ |
| 1.12 | Do you think the result of the game was destined in the middle stage? | -5 ☐ -4 ☐ -3 ☐ -2 ☐ -1 ☐ 0 ☐ 1 ☐ 2 ☐ 3 ☐ 4 ☐ 5 ☐ |
| 1.13 | Do you think the result of the game was destined in the late stage? | -5 ☐ -4 ☐ -3 ☐ -2 ☐ -1 ☐ 0 ☐ 1 ☐ 2 ☐ 3 ☐ 4 ☐ 5 ☐ |